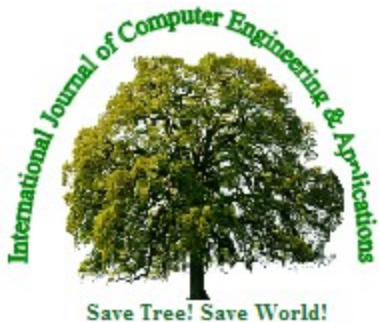

# A NEW WATERMARKING TECHNIQUE FOR SECURE DATABASE

Jun Ziang Pinn[1] and A. Fr. Zung[2]

[1,2]P. S. University for Technology, Harbin 150001, P. R. China

**ABSTRACT**

Digital multimedia watermarking technology was suggested in the last decade to embed copyright information in digital objects such images, audio and video. However, the increasing use of relational database systems in many real-life applications created an ever increasing need for watermarking database systems. As a result, watermarking relational database systems is now merging as a research area that deals with the legal issue of copyright protection of database systems. Approach: In this study, we proposed an efficient database watermarking algorithm based on inserting binary image watermarks in non-numeric mutli-word attributes of selected database tuples. Results: The algorithm is robust as it resists attempts to remove or degrade the embedded watermark and it is blind as it does not require the original database in order to extract the embedded watermark. Conclusion: Experimental results demonstrated blindness and the robustness of the algorithm against common database attacks.

**Keyword-** Watermarking, Database, Copyright Protection, Robustness,

## 1. INTRODUCTION

Security is of increasing concern with databases for database's high added values and extensive installation in modern information systems. In addition to encryption, watermarking techniques is practically proven as another possible solution to enhance databases' content security especially for copyright protection [1, 2, 3, 4, 5, 6] and data tampering detection [7]. Unlike encryption or hash description, typical watermarking techniques modify original data as a modulation of the watermark information, and inevitably cause permanent distortion to the original data, and therefore cannot meet the integrity requirement of the data in some applications. This underlying defect can be relieved by reversible watermarking techniques.

Most watermarking research concentrated on watermarking multimedia data objects such as still images and video and audio. However, watermarking of database systems started to receive attention because of the increasing use of database systems in many real-life applications.

Due to the different characteristics between images or audio and relational data, there exists no image or audio watermarking method suitable for watermarking relational databases. Therefore, relational database watermarking is, in fact, a process challenged by many factors such as data redundancy fewness, relational data out-of-order and frequent updating. Moreover, database systems watermarking have unique and sometimes complex, requirements that differ from those required for watermarking digital audio-visual products. Due to such unique requirements and challenges, literature on watermarking relational databases is very limited and has focused mainly on embedding short strings of binary bits in randomly selected locations in numerical databases.

## 2. RELATED WORK

Initially, most of the work on digital watermarking was concentrated on media like image, video, audio, VLSI design etc. However, in the recent years watermarking on databases started to receive attention. In general, the database watermarking techniques consist of two phases: *Watermark key* Embedding and Watermark key Verification. During embedding phase, a private key, original database act as inputs to watermark embedding algorithm. The watermarked database is then made publicly available. To verify the ownership of a suspicious database, the verification process is performed where the suspicious database and private key are the inputs for extraction algorithm. Finally, extraction algorithm display the result as suspicious database is original or not.

The idea to secure a database by digital watermarking technique was first coined by Khanna and Zane in 2000 [8]. In 2002, Agrawal et al. proposed a watermarking algorithm for relational databases that embeds the watermark key in the least significant bits (LSB) in selected attributes of a relation [9]. This technique does not provide a mechanism for multi-bit watermarks. For each row of a relation, a secure message authenticated code (MAC) is computed and finally embedded into the targeted least significant bits. Li et al. [10] in 2005 have presented a technique for fingerprinting relational data by extending Agrawal et al.'s watermarking scheme. Sion et al. in 2004 proposed a watermarking technique that embeds watermark key in the data statistics [11].

Above relevant works all assume that minor distortions caused to some attribute data can be tolerated to some specified precision grade. However some applications in which relational data are involved cannot tolerate any permanent distortions and data's integrity needs to be authenticated. To meet this requirement, we propose a reversible watermarking technique for lossless authentication of relational databases. Considering the typical case of randomly generated data sequence with even distribution as the host data, the scheme takes advantage of the uneven distribution of the error of two even-distributed variables and gains embedding capacity from reversible histogram expansion.

## 2.1. Proposed Algorithm

In our proposed algorithm, a binary image is used to watermark relational databases. The bits of the image are segmented into short binary strings that are encoded in non-numeric, multi-word attributes of selected tuples of the database. The embedding process of each short string is based on creating a double-space at a location determined by the decimal equivalent of the short string. Extraction of a short string is done by counting number of single-spaces between two separated doublespace locations. The image watermark is then constructed by converting the decimals into binary strings. A major advantage of using the space-based watermarking is the large bit-capacity available for hiding the watermark.

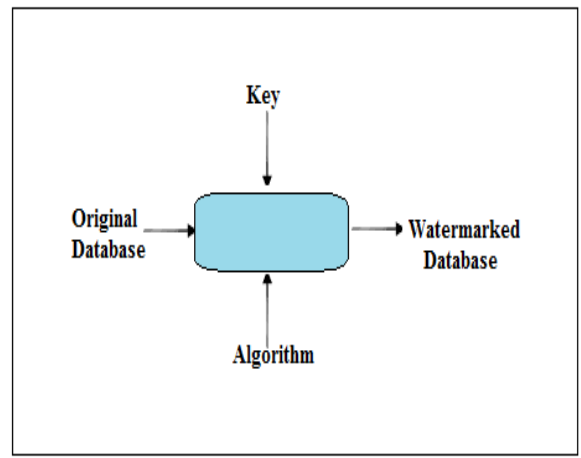

Fig.1 Watermark Embedding Process

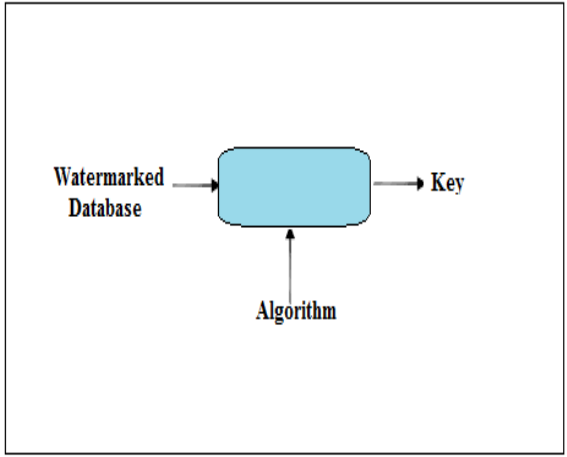

Fig.2 Watermark Extraction Process

Our proposed algorithm has two procedures: watermark embedding procedure and watermark extraction procedure. The two procedures are described below.

**Watermark embedding procedure:** The watermark embedding procedure consists of the following operational steps:

**Step 1:** Arrange the watermark image into m strings each of n bits length

**Step 2:** Divide the database logically into sub-sets of tuples. A sub-set has m tuples

**Step 3:** Embed the m short stings of the watermark image into each m-tuple sub-set

**Step 4:** Embed the n-bit binary string in the corresponding tuple of a sub-set as follows:

i. Find the decimal equivalent of the string. Let the decimal equivalent be d

ii. Embed the decimal number d in a pre-selected nonnumeric, multi-word attribute by creating a doublespace after d words of the attribute

**Step 5:** Repeat step 4 for each tuple in the subset **Step 6:** Repeat steps 4 and 5 for each subset of the database under watermarking

**Watermark extraction procedure:** The watermark embedding procedure consists of the following operational steps:

**Step 1:** Arrange the watermark image into m strings each of n bits length

**Step 2:** Divide the database logically into sub-sets of tuples. A sub-set has m tuples

**Step 3:** Embed the m short stings of the watermark image into each m-tuple sub-set.

**Step 4:** Embed the n-bit binary string in the corresponding tuple of a sub-set as follows:

i. Find the decimal equivalent of the string and give it the symbol d

ii. Embed the decimal number d in a pre-selected nonnumeric, multi-word attribute by creating a double space after d words of the attribute

**Step 5:** Repeat step 4 for each tuple in the subset

**Step 6:** Repeat steps 4 and 5 for each subset of the database under watermarking

## 3. RESULTS

In this section we present some experimental results to demonstrate the effectiveness of our proposed scheme. We ran experiments on MS SQL Server 2000 using .Net connectivity on a Windows 7 operating system with a core i3 processor, 2.0 G of memory, and 180-GB disk drive. The dynamic configure SQL Server memory is at most 1024 MB. The minimum query memory is 1024 KB. The database we used in the experiments was the transactional. Initially, we selected a table having 56,118 tuples and 22 attributes. Among 22 attributes we made 9 domains. After generating keys, we first tested the computational cost of watermark embedding and detection. Two experiments were performed. Each experiment was performed on the table which has 56,118 tuples. The average time required for watermark embedding was 1070 seconds. For the watermark verification, the required time is 256 seconds on average. These results indicate our algorithm performs well enough to be used in real-world applications.

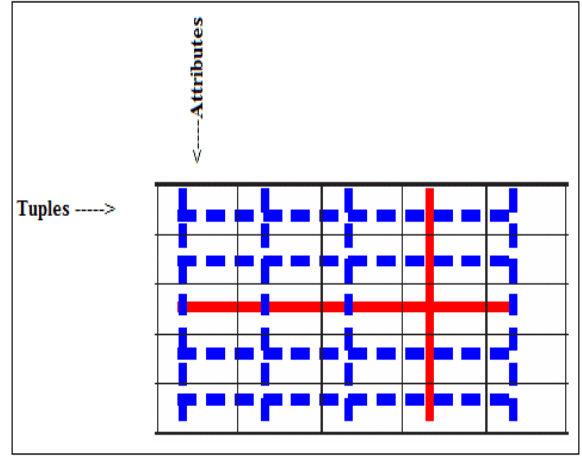

Fig.3 Database Schema

## 4. CONCLUSION

In this study, we proposed a watermarking algorithm based on hiding watermark bits in spaces of non-numeric, multi-word, attributes of subsets of tuples. A major advantage of using this approach is the large bit-capacity available to hide large watermarks.

The proposed technique must be suitable for different areas like, e-banking, multimedia industries, film industries etc.